\documentstyle[epsfig, color]{elsart}
\input amssym.def
\input amssym
\def\lesssim{\mathrel{\hbox{\rlap{\hbox{\lower4pt\hbox{$\sim$}}}\hbox{$<$}}}}
\begin{document}
\begin{frontmatter}
\title{Study of the $\eta$N scattering amplitude through the associated photoproduction of
$\phi$- and $\eta$-mesons
in the region of the
N*(1535) resonance}

\author{Matthias F. M. Lutz}
\address{GSI, Planckstrasse 1,  D-64291 Darmstadt, Germany}
\author{Madeleine Soyeur}
\address{DAPNIA/SPhN, CEA/Saclay, F-91191 Gif-sur-Yvette Cedex, France}

\begin{abstract}

The $\gamma p \rightarrow \phi \eta p$
reaction is studied in the kinematic region where the $\eta p$ final state
originates dominantly from the decay of the N*(1535) resonance. The threshold
laboratory photon energy for this reaction (at the peak of the S11 resonance) is $E_\gamma^{Lab}\,=\,$3 GeV.
We will discuss it somewhat above threshold, at $E_\gamma^{Lab}\simeq\,4-5$ GeV,
in order to reach lower (absolute) values of the squared 4-momentum transfer from the initial photon
to the final $\phi$-meson. In these conditions, we expect the t-channel $\pi^0$-
and $\eta$-meson exchanges to drive the dynamics underlying the $\gamma p \rightarrow \phi \eta p$
process.
The initial photon dissociates into the final $\phi$-meson and a virtual pseudoscalar
meson ($\pi^0$ or $\eta$). The virtual pseudoscalar meson
scatters from the proton target to produce the final $\eta p$ state.
The $\pi^0 p \rightarrow \eta p$ and $\eta p \rightarrow \eta p$
amplitudes are derived in the framework of a coupled-channel
effective field theory of meson-baryon scattering.
We found the
$\eta$-meson exchange to be largely dominant. The $\eta$-$\pi^0$ interference is of the order
of $20\,-\,30 \%$. The sign of this term is not known and has a significant
influence on the results. The $\pi  N\rightarrow \eta N$ amplitude being largely determined by
data on the $\pi^- p \rightarrow \eta \, n$ reaction,
we found that the $\gamma p \rightarrow \phi \eta p$ reaction cross section is rather directly related to the
$\eta$-nucleon scattering amplitude in the N*(1535) resonance region. Accurate data on the
$\gamma p \rightarrow \phi \eta p$ process would therefore
put additional constraints on this still poorly known amplitude.

\vskip 0.3truecm


\end{abstract}

\end{frontmatter}

\newpage
\section{Introduction}

The structure of the $\eta$-nucleon scattering amplitude close to threshold
($\sqrt s$ = 1.486 GeV) is of much interest because it appears dominated by
the presence of a baryon resonance slightly above threshold, the N*$_{1/2^-}$(1535).
The width of the N*(1535) is of the order of 150 MeV. Its two main decay channels
are the N$\pi$ (35-55 $\%$) and the N$\eta$ (30-55 $\%$) final states \cite{PDG}.

The $\eta$-nucleon scattering length, $a_{\eta N}$, characterizes the behaviour
of the $\eta$-nucleon interaction at threshold. It is a complex quantity whose
real part is still poorly determined. We refer to the recent work of Green and
Wycech \cite{GreenWycech} for a discussion of the range of values found in the literature
(from $\sim$ 0.3 to $\sim$1 fm).
The main reason for the spread in values is that $a_{\eta N}$ is obtained
indirectly and through model-dependent analyses of pion- and photon-induced
$\eta$ production reactions. The $\eta$-nucleon scattering length is a key quantity
to assess the possibility of forming $\eta$-nuclear quasi-bound states \cite{Niskanen}.

In this work, we propose to study the threshold behaviour of the $\eta$-nucleon scattering
amplitude through the $\gamma \, p \rightarrow \phi \, \eta \, p$ reaction
in the particular kinematics where the invariant mass of the $\eta \, p$
pair is close to the N*(1535) mass. We tune the incident laboratory photon energy in order
to reach low momentum transfers, i.e. a sufficiently small $|t_{min}|$.
At the $\phi$ threshold, $E_\gamma^{Lab}$=3 GeV and $|t_{min}|$ is 1.2 GeV$^2$.
We will consider values of $E_\gamma^{Lab}$ ranging from 4 to 5 GeV.
At $E_\gamma^{Lab}$=4 GeV, $|t_{min}|$ is 0.38 GeV$^2$.
At 5 GeV, it is 0.26 GeV$^2$.
Our main argument is that the $\gamma \, p \rightarrow \phi \, \eta \, p$ process
in these kinematics ($|t|<$ 1 GeV$^2$) is dominated by the $\eta$ t-channel exchange and
offers the possibility to test the $\eta$-nucleon scattering amplitude
close to threshold. Both $\pi$- and $\eta$-exchanges can contribute.
The dominance of the $\eta$-exchange
in the t-channel
comes mainly from the property that
the $\phi$-meson radiative decay probability to the $\eta \gamma$ channel
is an order of magnitude larger than to the
$\pi \gamma$ channel (despite the larger phase space available for the latter decay).
This is ultimately related to the large $s\bar s$ content of the $\eta$-meson.
The s-wave $\pi \, p \rightarrow \eta \, p$ amplitude is also significantly
smaller than the $\eta \, p \rightarrow \eta \, p$ amplitude.

Our results on the $\gamma \, p \rightarrow \phi \, \eta \, p$ reaction
are based on $\pi \, p \rightarrow \eta \, p$ and $\eta \, p \rightarrow \eta \, p$ scattering
amplitudes obtained in the unitary coupled-channel model of Ref. \cite{Lutz1}.
These amplitudes reproduce a large set of pion-nucleon and photon-nucleon
scatte\-ring data in the energy range $1.4<\sqrt s <1.8$ GeV, in particular
the pion- and photon-induced $\eta$-meson production cross sections.
The value obtained for $a_{\eta N}$ is (1.03 + i 0.49) fm \cite{Lutz1}, in close agreement
with the findings of Ref. \cite{GreenWycech}.

Our t-channel calculation of the $\gamma \, p \rightarrow \phi \, \eta \, p$ reaction cross section
in the N*(1535) region is described in Section 2. We discuss the $\eta$- and $\pi^0$-exchanges
and their interference. The latter is constructive or destructive depending on the relative
sign of the couplings constants $g_{\phi\pi\gamma}$ and $g_{\phi\eta\gamma}$ of the
corresponding anomalous interaction Lagrangians.
We display numerical results for the $\gamma \, p \rightarrow \phi \, \eta \, p$ reaction cross section
in Section 3. We show the expected t-distributions at $E_\gamma^{Lab}$=4 GeV and 5 GeV
and emphasize the role of the
double $\eta$-pole term and of the $\eta$-$\pi^0$ interference in these quantities.
A few concluding remarks are given in Section 4.

\section{The $\gamma \, p \rightarrow \phi \, \eta \, p$ reaction cross section
in the N*(1535) region}
The t-channel $\pi$- and $\eta$-exchange amplitudes contributing to the
$\gamma \, p \rightarrow \phi \, \eta \, p$
process
in the N*(1535) region are shown in Figs. 1 and 2.
\vglue 0.4true cm
\begin{figure}[h]
\noindent
\begin{center}
\mbox{\epsfig{file=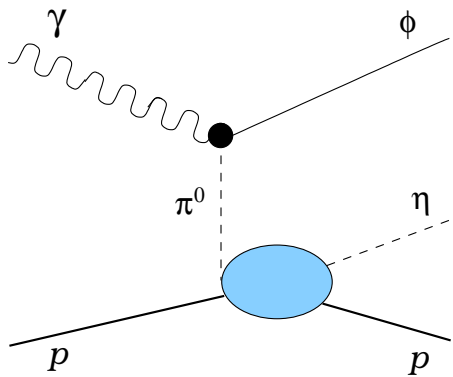, height=5 truecm}}
\end{center}
\vskip 0.4 true cm
\caption{Pion-exchange contribution to the $\gamma \, p \rightarrow \phi \, \eta \, p$ process.}
\label{pionamplitude}
\end{figure}
\vglue 0.4truecm
\vglue 0.4true cm
\begin{figure}[h]
\noindent
\begin{center}
\mbox{\epsfig{file=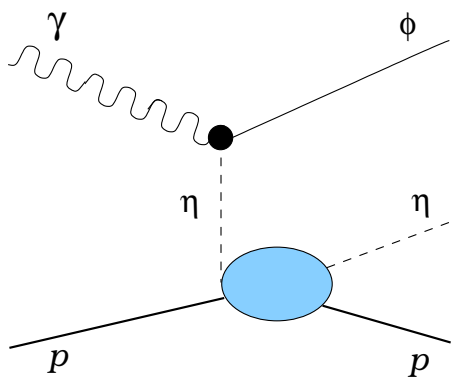, height=5 truecm}}
\end{center}
\vskip 0.4 true cm
\caption{$\eta$-exchange contribution to the $\gamma \, p \rightarrow \phi \, \eta \, p$ process.}
\label{etaamplitude}
\end{figure}
\vglue 0.4truecm
We calculate the  cross section for the $\gamma \, p \rightarrow \phi \, \eta \, p$ reaction,
assuming it is driven by the mechanisms displayed in Figs. 1 and 2. We have three
contributions, associated with the $\pi$-exchange, the $\eta$-exchange and their
interference respectively. The t-channel dominance is clearly an hypothesis which can but be
confirmed by measurements of differential cross sections. We note that recent data
on the $\gamma \, p \rightarrow \phi \, p$ reaction at $E_\gamma^{Lab}$=2.4 GeV \cite{Barth,Mibe}
show that this process is dominated by t-channel exchanges from threshold onwards.
The exact nature of these exchanges is however not clear \cite{Barth,Mibe}. It is
therefore of interest to study a $\phi$-photoproduction process in which
the excitation of the target to the N*$_{1/2^-}$(1535) is expected to favour specifically
unnatural parity exchanges, i.e. the $\pi$ and $\eta$ t-channel contributions.

The 4-momenta
of the photon, the proton, the $\phi$, the $\eta$ and the final proton are denoted
by $q$, $p$, $\bar q_{\phi}$, $\bar q_\eta$ and $\bar p$.
The photon, initial proton and final proton polarizations are indicated by the
symbols $\lambda_\gamma$, $\lambda$ and $\bar
\lambda$. The total
cross section reads
\begin{eqnarray}
\sigma_{\gamma \,p\rightarrow \phi \, \eta \,p}&=&
\frac{1}{|{\vec v}_\gamma-{\vec v}_p|}\,\frac{1}{2\,q^0}\,\frac{m_p}{p^0}
\int \frac{d^3 \vec{\bar q}_\phi}{(2\pi)^3}\,\frac{1}{2\,{\bar q}\,^0_\phi}\,
\int \frac{d^3 \vec{\bar q}_\eta}{(2\pi)^3}\,\frac{1}{2\,{\bar q}\,^0_\eta}\,
\int \frac{d^3 \vec{\bar p}}{(2\pi)^3}\,\frac{m_p}{{\bar p}\,^0}\,
\nonumber\\
&&\times(2\,\pi)^4\,\delta^4(q+p-{\bar q}_\phi-\bar q_\eta-\bar p)\sum_{\lambda_\gamma,
\lambda , \bar \lambda_\phi, \bar
\lambda }\frac{1}{4}\,|M_{\gamma \,p\rightarrow \phi\, \eta \,p}|^2.
\label{crosssection}
\end{eqnarray}

The photon-$\phi$-pseudoscalar meson vertices are described by the anomalous
interaction Lagrangian
\begin{eqnarray}
{\cal L}^{int}_{\phi \chi \gamma} =
e \,\frac{g_{\phi \chi \gamma}} {2 m_\phi} \, \varepsilon^{
\mu\nu\alpha\beta} \, \phi_\mu \, (\partial_\nu \chi)\, F_{\alpha \beta},
\label{anomalousLintpi}
\end{eqnarray}
where
\begin{eqnarray}
F_{\alpha \beta}\, =\partial_\alpha A_\beta \,-\,\partial_\beta A_\alpha
\label{defF}
\end{eqnarray}
\noindent
and $\chi$ denotes the chiral pseudoscalar mesons of interest, the pion or the $\eta$-meson.
Using this interaction Lagrangian to calculate the $\phi\rightarrow \pi^0\gamma$
and $\phi\rightarrow \eta\gamma$ partial widths, we have
\begin{eqnarray}
\Gamma_{\phi\rightarrow \chi\gamma}\, =\, g^2_{\phi \chi \gamma} \, \frac{\alpha} {24}
m_\phi (1- \frac {m_\chi^2}{m_\phi^2})^{3}.
\label{widthMgamma}
\end{eqnarray}
Identifying this expression with the experimental decay widths \cite{PDG}
\begin{eqnarray}
\Gamma_{\phi\pi^0\gamma}\, =\, (5.24 \, \pm \, 0.49) \, keV,
\label{widthpigamma}
\end{eqnarray}
\begin{eqnarray}
\Gamma_{\phi\eta\gamma}\, =\, (55.17 \, \pm \, 1.71) \, keV,
\label{widthetagamma}
\end{eqnarray}
we obtain
\begin{eqnarray}
|g_{\phi\pi\gamma}|\, \simeq\, 0.13,\mkern 30 mu |g_{\phi\eta\gamma}|\, \simeq\, 0.70.
\label{coupling constants}
\end{eqnarray}

If we want to compute the interference between the $\pi$- and $\eta$-exchanges,
we should know the relative sign of the coupling constants
$g_{\phi\pi\gamma}$ and $g_{\phi\eta\gamma}$. It is generally assumed
that the $\phi\rightarrow\pi\gamma$ transition is dominated by the
$\phi-\omega$ mixing. The $\phi-\omega$ mixing angle is not easy to determine because it
depends on the overlap of two narrow resonances far apart from each other.
It is therefore model-dependent and sensitive to the assumptions used to calculate
the $\omega$-meson self-energy far from the $\omega$ pole.
The $\phi\rightarrow\eta\gamma$ transition depends on the $\eta$-$\eta'$ mixing,
a phenomenon closely connected with the U(1) anomaly of QCD and still
under analysis in
different schemes \cite{Escribano}.
Based on recent analyses of the available data, it is most often
found that both $g_{\phi\pi\gamma}$ and $g_{\phi\eta\gamma}$
have the same sign as $g_{\omega\pi\gamma}$, and hence a positive
relative sign \cite{Bramon,Titov}. The opposite
conclusion has also been reached \cite{Benayoun}. In view of the uncertainties
in the SU(3) symmetry breaking mechanisms responsible for the
mixing angles, we consider this relative sign as yet undetermined.

We will now establish the dominance of the $\eta$-exchange process
in the $\gamma \, p \rightarrow \phi \, \eta \, p$ reaction in the
kinematics of interest ($E_\gamma^{Lab}\backsimeq$ 4-5 GeV and  $|t_{min}|<|t| \lesssim$ 1 GeV$^2$).
Using the interaction
Lagrangian (2), the squared amplitudes corresponding to the $\pi$-exchange,
the $\eta$-exchange and their interference are
\begin{eqnarray}
\sum_{\lambda_\gamma, \lambda ,\bar
\lambda_\phi, \bar \lambda }\,\frac{1}{4}\,\mid M_{\gamma \,p\rightarrow \phi \, \eta \,p}^{\pi-exchange}\mid^2
&=&\frac {e^2\,g_{\phi\pi\gamma}^2} {4\,m_{\phi}^2}\frac{(m_\phi^2-t)^2}{(t-m_\pi^2)^2}\,\frac{1}{2}\,
\sum_{\lambda ,\bar
\lambda }\,|M_{\pi\,p \to \eta\,p}|^2,
\label{piexchange}
\end{eqnarray}
\begin{eqnarray}
\sum_{\lambda_\gamma, \lambda ,\bar
\lambda_\phi, \bar \lambda }\,\frac{1}{4}\,\mid M_{\gamma \,p\rightarrow \phi \, \eta \,p}^{\eta-exchange}\mid^2
&=&\frac {e^2\,g_{\phi\eta\gamma}^2} {4\,m_{\phi}^2}\frac{(m_\phi^2-t)^2}{(t-m_\eta^2)^2}\,\frac{1}{2}\,
\sum_{\lambda ,\bar
\lambda }\,|M_{\eta\,p \to \eta\,p}|^2,
\label{etaexchange}
\end{eqnarray}
\begin{eqnarray}
\sum_{\lambda_\gamma, \lambda ,\bar
\lambda_\phi, \bar \lambda }\,\frac{1}{4}\,\mid M_{\gamma \,p\rightarrow \phi \, \eta \,p}^{interference}\mid^2
=\frac {e^2\,g_{\phi\pi\gamma}\,g_{\phi\eta\gamma}} {4\,m_{\phi}^2}\frac{(m_\phi^2-t)^2}{(t-m_\pi^2)(t-m_\eta^2)}
\nonumber
\\\frac{1}{2}\,
\sum_{\lambda ,\bar
\lambda }\,(M^+_{\pi\,p \to \eta\,p}M_{\eta\,p \to \eta\,p}\,+\,M^+_{\eta\,p \to \eta\,p}M_{\pi\,p \to \eta\,p}).
\label{interference}
\end{eqnarray}
We work in the photon-proton center of mass reference frame
where the total energy of the reaction is denoted by $\sqrt s$.
In that reference frame, the photon 3-momentum is -${\vec p}$
and the 3-momentum of the $\eta\,p$ pair is -${\vec{\bar q}_\phi}$.
We define the invariant mass $\sqrt{{\bar w}^2}$ of the final $\eta \,p$ pair by
\begin{eqnarray}
{\bar w}^2 = (p+q-{\bar q}_\phi)^2
 = s+ m_\phi^2- 2\,\sqrt{s}\,\sqrt{m_\phi^2+{\vec{\bar q}_\phi}^{\,2}} \,
\label{eqkin1}
\end{eqnarray}
and express the 4-momentum transfer $t=(q-{\bar q}_\phi)^2$ as function of
that variable,
\begin{eqnarray}
t\;(s,{\bar w}^2,\cos \theta) &=&
 m_\phi^2 -\frac{1} {2s} (s-m_p^2)\, (s+m_\phi^2-{\bar w}^2)\,\nonumber\\
&&\mkern 110 mu \times \Big(1 - \sqrt{1-\frac{4\,m_\phi^2\,s}{(s+m_\phi^2-{\bar w}^2)^2}} \,\cos \theta\Big),
\label{eqkin2}
\end{eqnarray}
where $\theta$ is the angle between the initial photon and the produced $\phi$-meson.
We define the notation
$t_-(s,{\bar w}^2)\equiv t_{min}(s,{\bar w}^2)=t\;(s,{\bar w}^2,\cos \theta=+1)$ and
$t_+(s,{\bar w}^2)=t\;(s,{\bar w}^2,\cos \theta =-1)$.

Using the above variables, the differential cross section for the
${\gamma \,p\rightarrow \phi \, \eta \,p}$ reaction with respect to $t$ and ${\bar w}^2$
is given by
\begin{eqnarray}
\frac{d\sigma_{\gamma \,p\rightarrow \phi \, \eta \,p}} {dt\, d{\bar w}^2}&=&
 \frac{\alpha\,m_p} {16 \, m_{\phi}^2\, s\,|\vec{p}\,|^2}
\int \frac{d^3 \vec{\bar q}_\eta}{(2\pi)^3}\,\frac{1}{2\,{\bar q}\,^0_\eta}\,
\int \frac{d^3 \vec{\bar p}}{(2\pi)^3}\,\frac{m_p}{{\bar p}\,^0}
\nonumber\\
&&\times (2\,\pi)^4
 \delta^4(q+p-{\bar q}_\phi-\bar q_\eta-\bar p)
\nonumber\\
&&\times
\{\frac{g_{\phi \pi \gamma}^2} {4\pi} \,\frac{(m_\phi^2-t)^2}{(t-m_\pi^2)^2}
 \sum_{\lambda,
\bar\lambda }\frac{1}{2}\,|M_{\pi \,p\rightarrow \eta \,p}|^2
\nonumber\\
&&+ \frac{g_{\phi \eta \gamma}^2} {4\pi} \,\frac{(m_\phi^2-t)^2}{(t-m_\eta^2)^2}
 \sum_{\lambda,
\bar\lambda }\frac{1}{2}\,|M_{\eta \,p\rightarrow \eta \,p}|^2
\nonumber\\
&&+ \frac{g_{\phi \pi \gamma}\, g_{\phi \eta \gamma}} {4\pi} \,\frac{(m_\phi^2-t)^2}{(t-m_\pi^2)(t-m_\eta^2)}
\nonumber\\
&&\times
\frac{1}{2}\,
\sum_{\lambda ,\bar
\lambda }\,(M^+_{\pi\,p \to \eta\,p}M_{\eta\,p \to \eta\,p}\,+\,M^+_{\eta\,p \to \eta\,p}M_{\pi\,p \to \eta\,p})\}.
\label{crosssectdiff}
\end{eqnarray}
\noindent The total cross section is obtained by integrating
over $t$ from $t_+$ to $t_-$ and over ${\bar w}^2$ from $(m_p+m_\eta)^2$ to
$(\sqrt{s}-m_\phi)^2$. In all the results presented in the next section, we will
restrict the interval of values of $t$ from $t \,=\, -1$ GeV$^2$ to $t_{min}$,
the expected range of validity of our model. Because $t_+$ extends to much
lower values (for $\bar w =\,$1.54 GeV, $t_+ \, = \,-5.8$ GeV$^2$ at
$E_\gamma^{Lab}=\,$5 GeV for example), we will refrain from displaying integrated cross sections.

We note that Eq. (\ref{crosssectdiff}) is gauge-invariant because of the specific form of the
interaction Lagrangian (\ref{anomalousLintpi}).

It is of interest to study the pole structure of the three contributions to the cross section
($\pi^0$-exchange, $\eta$-exchange, interference) as it determines the shape of the
t-distributions.
We have the following decompositions:

\begin{eqnarray}
&& \frac{(m_\phi^2-t)^2}{(t-m_\pi^2)^2} =
\frac{(m_\phi^2-m_\pi^2)^2}{(t-m_\pi^2)^2}
- 2\,\frac{m_\phi^2-m_\pi^2}{t-m_\pi^2} + 1,
\\
&& \frac{(m_\phi^2-t)^2}{(t-m_\eta^2)^2} =
\frac{(m_\phi^2-m_\eta^2)^2}{(t-m_\eta^2)^2}
- 2\,\frac{m_\phi^2-m_\eta^2}{t-m_\eta^2} + 1,
\\
&& \frac{(m_\phi^2-t)^2}{(t-m_\eta^2)\,(t-m_\pi^2)} =
\frac{(m_\phi^2-m_\eta^2)^2}{t-m_\eta^2}\,\frac{1}{m_\eta^2-m_\pi^2}
\nonumber\\
&& \qquad \qquad \qquad \quad\quad
-\frac{(m_\phi^2-m_\pi^2)^2}{t-m_\pi^2}
\,\frac{1}{m_\eta^2-m_\pi^2} + 1.
\label{polestructure}
\end{eqnarray}

Simple effects can be understood from these expressions irrespectively
of the dynamical aspects of the scattering amplitudes and  of the strength of
the $\gamma \chi \phi$ vertex.
It is easy to see from Eq. (14) that the double pion pole term
(proportional to $\frac{1}{(t-m_\pi^2)^2}$)
dominates the pion-exchange contribution close
to $t_{min}$, becomes comparable to the single pion pole term (proportional to $\frac{1}{(t-m_\pi^2)}$)
around $t=-0.5$ GeV$^2$ and smaller than the latter for larger values of $\mid t \mid$. In the case of the
$\eta$-exchange displayed in Eq. (15),
the single $\eta$ pole term is always dominant in the kinematic range under consi\-deration. Close to $t_{min}$,
the double $\eta$ pole term represents typically  25-30$\%$ of the total $\eta$-exchange contribution.
It is nevertheless instrumental in producing a rather sharp drop in the differential cross section.
In the interference term (16), the t-dependence is largely given by the single pion pole term at low $\mid t \mid$.

To derive Eq. (13), we have assumed that the $\gamma \eta \phi$ transition form factor is one.
We have no information on that quantity but it influences significantly the outcome of
our calculation. If the form factor is hard, its effect will be rather small. If it is
soft, it could affect substantially the t-dependence of the ${\gamma \,p\rightarrow \phi \, \eta \,p}$ reaction.
The only argument we see in favour of a hard form factor is that there is no obvious
intermediate state to build a form factor in the $\eta$ direction. The $\eta$-meson couples
dominantly to two photons and to three pions. To construct a form factor in the $\eta$-channel,
one would need a significant decay of the $\phi$-meson into a photon and two vector particles
or into a photon and three pions. The only available information is an upper limit of 5 10$^{-4}$ on
the branching ratio of the $\phi$-meson to the $\rho \gamma \gamma$ channel \cite{PDG}.
In the absence of more significant data, it is not possible to gain a reasonable understanding of
the $\gamma \eta \phi$ form factor for space-like $\eta$-mesons. We will therefore set it to one,
keeping this assumption in mind.

Finally, if our model takes into account the $\eta$-nucleon final state interaction to all orders,
it does not treat $\phi$-nucleon rescattering in the outgoing channel.
We do not expect this rescattering to be very important in the kinematics under
consideration, i.e. with a large relative momentum between the
$\phi$-meson emitted at small angles and the recoiling target products.
In particular, we are not in the threshold regime where cryptoexotic $B_\phi$
baryons or $\phi N$ resonances
could enlarge final state interactions \cite{Sibirtsev}.

\section{Numerical results for the $\gamma \,p\rightarrow \phi \, \eta \,p$ reaction}
We proceed to the calculation of the $\gamma \,p\rightarrow \phi \, \eta \,p$
cross section as outlined in the previous section using the
$\pi \, p \rightarrow \eta \, p$ and $\eta \, p \rightarrow \eta \, p$ scattering
amplitudes obtained in the model of Ref. \cite{Lutz1}.
These amplitudes are displayed in Fig. 3 and 4.
\vglue 1 true cm
\begin{figure}[h]
\noindent
\begin{center}
\mbox{\epsfig{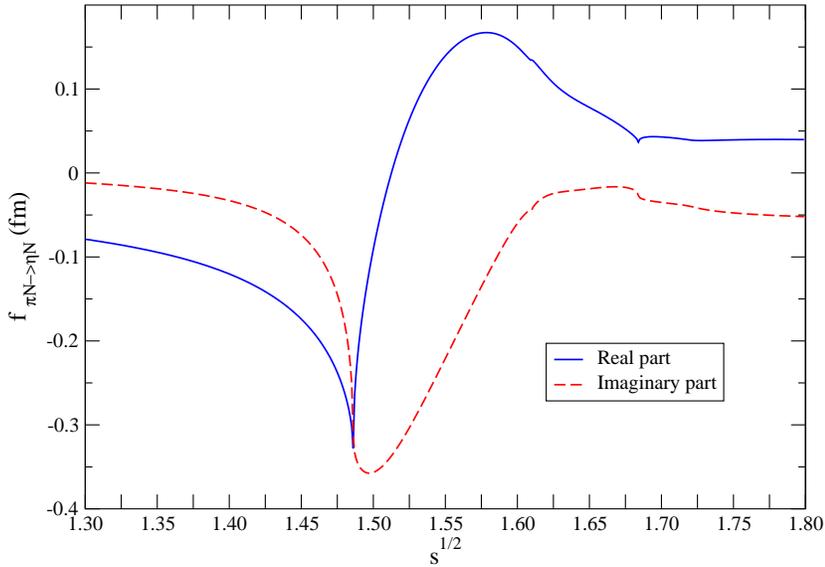}}
\end{center}
\vskip 0.4 true cm
\caption{Real and imaginary parts of the s-wave $\pi$N$\rightarrow\eta$N scattering amplitude (from Ref. \cite{Lutz1}).}
\label{pietaNamplitude}
\end{figure}
\vglue 0.7truecm
\vskip 0.4true cm
\begin{figure}[h]
\noindent
\begin{center}
\mbox{\epsfig{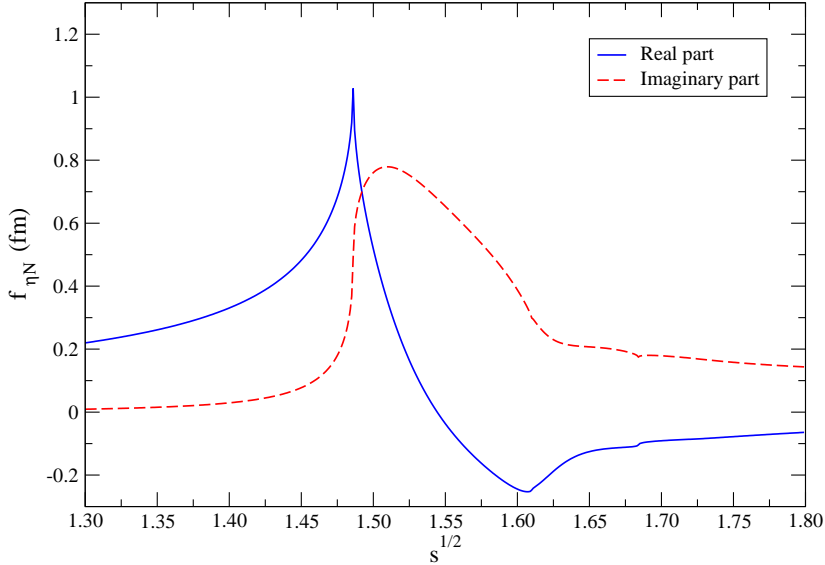}}
\end{center}
\vskip 0.4 true cm
\caption{Real and imaginary parts of the s-wave $\eta$N scattering amplitude (from Ref. \cite{Lutz1}).}
\label{etaNamplitude}
\end{figure}
We notice that the $\pi$N$\rightarrow\eta$N amplitude is about two to three times smaller
than the $\eta$N scattering amplitude in the region of the N*(1535) resonance.
Combined with the very unfavourable ratio
$|{g_{\phi\pi\gamma}|^2}/{|g_{\phi\eta\gamma}|^2}\thickapprox \,$1/29 of the ano\-malous
coupling constants, this effect suggests that the pion contribution to the
$\gamma \,p\rightarrow \phi \, \eta \,p$ cross section will be much smaller, typically by two
orders of magnitude, than the $\eta$ contribution.

It is interesting to remark that the amplitude displayed in Fig. 4 is very similar
to the corresponding quantity obtained in Ref. \cite{GreenWycech}, suggesting
that our calculation does not depend too much on the specific model used to derive
the amplitudes. We emphasize also that the real part of the $\eta$N scattering amplitude is
mostly visible in a narrow band of total center of mass energies, i.e. from threshold until
1.5 GeV. Beyond $\sqrt s$=1.52, the cross section will be dominated by the imaginary
part of the amplitude.

Consequently we will show results for two values of the invariant mass of the $\eta$N pair,
$\bar w$=1.49 GeV (very close to threshold) and  $\bar w$=1.54 GeV (at the resonance
peak).

We consider first $\bar w$=1.49 GeV and display
${d\sigma_{\gamma \,p\rightarrow \phi \, \eta \,p}}/ {dt \,d{\bar w}}$
at $E_\gamma^{Lab}=\,$ 4 GeV and $E_\gamma^{Lab}=\,$ 5 GeV. This is shown
in Figs. 5 and 6 respectively.

\begin{figure}[h]
\noindent
\begin{center}
\mbox{\epsfig{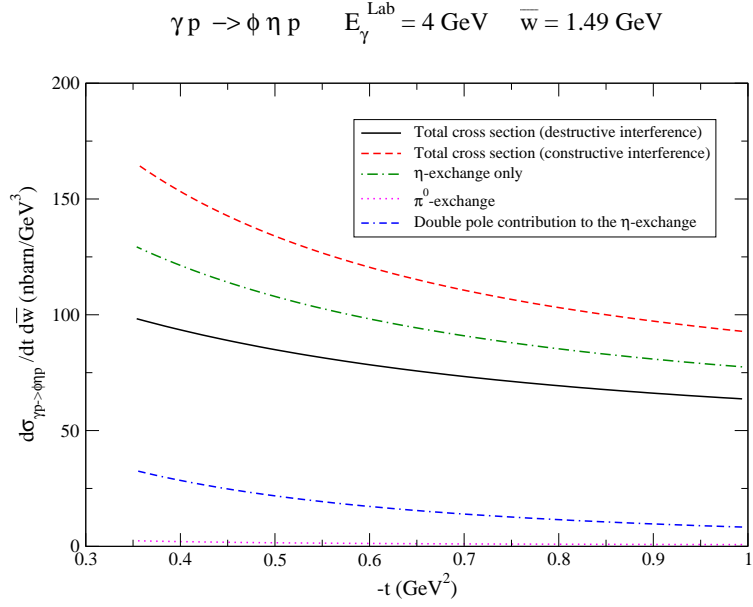}}
\end{center}
\vskip 0.4 true cm
\caption{Differential cross section
${d\sigma_{\gamma \,p\rightarrow \phi \, \eta \,p}}/ {dt \,d{\bar w}}$
at $E_\gamma^{Lab}=\,$ 4 GeV  for a total center of mass energy of the $\eta$N pair of 1.49 GeV.
The full and dashed lines are the total differential cross sections assuming a destructive and a constructive
$\pi^0$-$\eta$ interference respectively. The dot-dashed line is the full contribution of the
$\eta$-meson exchange while the dot-double-dashed line shows the double $\eta$-pole contribution.
The dotted line is the $\pi^0$-exchange contribution.}
\label{R4}
\end{figure}
\begin{figure}[h]
\noindent
\begin{center}
\mbox{\epsfig{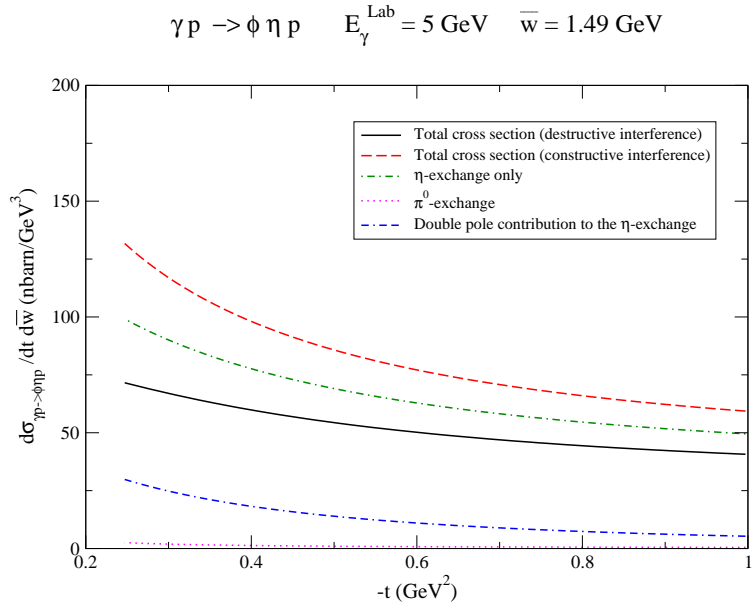}}
\end{center}
\vskip 0.4 true cm
\caption{Same as Fig. 5 at 5 GeV.}
\label{R45}
\end{figure}
As discussed earlier, we do not know the relative sign of the
${g_{\phi\pi\gamma}}$ and ${g_{\phi\eta\gamma}}$ coupling constants.
We have therefore considered both signs. If the coupling constants
are of the same sign, the $\eta-\pi^0$ interference is destructive
because of the opposite sign of the amplitudes (solid curve). If
${g_{\phi\pi\gamma}}$ and ${g_{\phi\eta\gamma}}$ are of opposite sign,
the $\eta-\pi^0$ interference is constructive (dashed curve).
In order to evaluate the importance of this interference, we
show in addition the $\eta$-exchange contribution (dot-dashed line) and the
$\pi^0$-exchange contribution (dotted line). Close to $t_{min}$, the
$\eta-\pi^0$ interference suppresses or enhances the differential
cross section by about 30$\%$. As anticipated, the $\pi^0$-exchange contribution
is completely negligible. We plot also the double pole contribution to
the $\eta$-exchange (dot-double-dashed line). It represents typically a quarter of the $\eta$-exchange term
but contributes very significantly to the t-dependence of the differential cross section.

If the $\eta-\pi^0$ interference is constructive, the pole structure of the $\eta$-exchange
and of the interference
leads to a rather sharp t-dependence close to $t_{min}$. This effect increases with increasing
laboratory photon energy as a lower $|t_{min}|$ can be reached. If the $\eta-\pi^0$ interference is
destructive, the terms driving the increase of the differential cross section
at low $|t|$ cancel significantly, leading to a rather flat behaviour.

We
show in Figs. 7 and 8 the t-distributions for the total $\eta$N center of mass energy
$\bar w$ = 1.54 GeV, taken to be close to the N*(1535) mass. The comparison to the
results obtained at $\bar w$ = 1.49 GeV
for the same value of the incident photon laboratory energy shows that the cross section increases.
This is a consequence of the opening of the $\eta$N phase space and of the presence of the
N*(1535) resonance (implying a large imaginary part in the amplitude). Otherwise the features
of the differential cross section
${d\sigma_{\gamma \,p\rightarrow \phi \, \eta \,p}}/ {dt \,d{\bar w}}$ are very similar
for both values of $\bar w$.
\begin{figure}[h]
\noindent
\begin{center}
\mbox{\epsfig{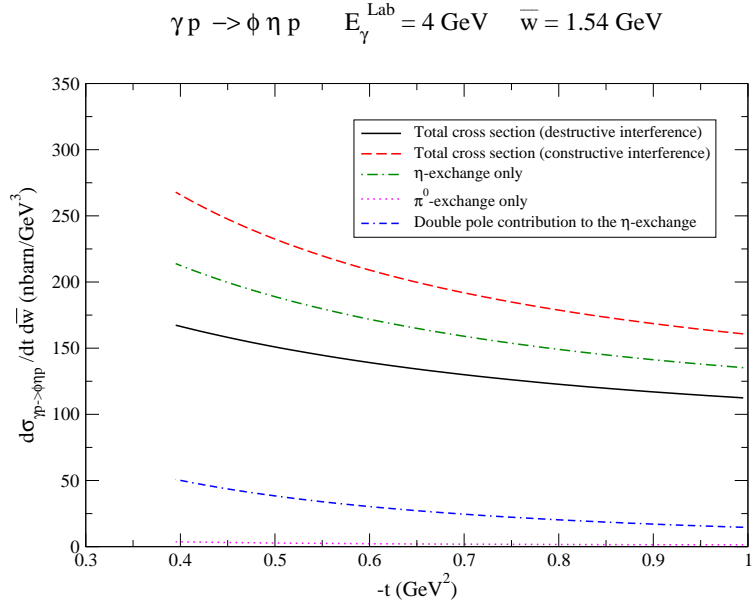}}
\end{center}
\caption{Differential cross section
${d\sigma_{\gamma \,p\rightarrow \phi \, \eta \,p}}/ {dt \,d{\bar w}}$
at $E_\gamma^{Lab}=\,$ 4 GeV for a total center of mass energy of the $\eta$N pair of 1.54 GeV.
The full and dashed lines are the total differential cross sections assuming a destructive and a constructive
$\pi^0$-$\eta$ interference respectively. The dot-dashed line is the full contribution of the
$\eta$-meson exchange while the dot-double-dashed line shows the double $\eta$-pole contribution.
The dotted line is the $\pi^0$-exchange contribution.}
\label{R5}
\end{figure}
\begin{figure}[h]
\noindent
\begin{center}
\mbox{\epsfig{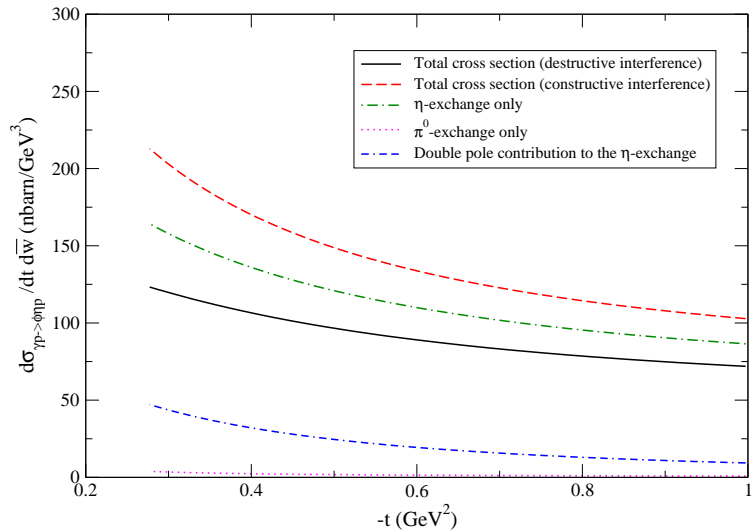}}
\end{center}
\caption{Same as Fig. 7 at $E_\gamma^{Lab}=\,$ 5 GeV.}
\label{F8}
\end{figure}

We emphazise however that the differential cross section
${d\sigma_{\gamma \,p\rightarrow \phi \, \eta \,p}}/ {dt \,d{\bar w}}$ at $\bar w$ = 1.49 GeV and at
$\bar w$ = 1.54 GeV tests different parts of the $\eta$N scattering amplitudes, the real and imaginary
parts at threshold on the one hand and the imaginary part at the resonance peak on the other hand.
We remark also that the $\pi N \rightarrow \eta N$ scattering amplitude is very much constrained
by data on the $\pi^- p \rightarrow \eta n$ cross section close to threshold \cite{GreenWycech,Lutz1}.
If our model is correct,
accurate data on the ${d\sigma_{\gamma \,p\rightarrow \phi \, \eta \,p}}/ {dt \,d{\bar w}}$ reaction
at low $|t|$ could therefore be interpreted in terms of the $\eta N$ scattering amplitude. We have underlined two
uncertainties in such an analysis, the relative sign of the ${g_{\phi\pi\gamma}}$ and ${g_{\phi\eta\gamma}}$
coupling constants and a possible effect due to a $\gamma \eta \phi$ transition form factor for space-like
$\eta$'s.

\section{Conclusion}

We have studied the $\gamma \,p\rightarrow \phi \, \eta \,p$ reaction
with the idea of using future accurate data on this process to gain
understanding of the threshold behaviour of the $\eta$-nucleon scattering amplitude.
We chose kinematic conditions where a t-channel meson-exchange description is expected
to be valid (i.e. low momentum transfers). We showed that the process in
which the initial photon dissociates into a $\phi$-meson and a virtual $\eta$-meson
scattering from the proton target to produce an $\eta$ proton final state is dominant
for ($\eta\,$p) pairs with invariant masses close to the N*(1535) mass.
It can however interfere with the analogous process where a $\pi^0$-meson
is exchanged. The $\pi^0$-exchange term is negligible.
The sign of the $\eta-\pi^0$ interference is at present not known. We show how it changes
both the t-dependence and the absolute value of the differential cross section.

In view of the fact that the $\pi N \rightarrow \eta N$ scattering amplitude
in the N*(1535) resonance region is rather well-known, accurate data on the
$\gamma \,p\rightarrow \phi \, \eta \,p$ reaction at $E_\gamma^{Lab}=\,$ 4-5 GeV
and low t would put additional
constraints on the still poorly controlled $\eta$-proton scattering amplitude
close to threshold. The expected cross sections are small but do not appear out
of reach of present experimental facilities.

\section{Acknowledgement}

One of us (M.S) is indebted to Slawomir Wycech for a very useful discussion
and to the hospitality of the GSI Theory Group, where part
of this work was done. We are thankful to our referees for comments
which led us to significantly improve the contents of this paper.
We acknowledge the support of the European Community-Research Infrastructure Activity under
the FP6 "Structuring the European Research Area" programme
(Hadron Physics, contract number RII3-CT-2004-506078).

\vglue 0.4truecm

\end{document}